\documentclass[sn-mathphys,Numbered]{sn-jnl}

\usepackage{graphicx}%
\usepackage{multirow}%
\usepackage{amsmath,amssymb,amsfonts}%
\usepackage{amsthm}%
\usepackage{mathrsfs}%
\usepackage[title]{appendix}%
\usepackage{xcolor}%
\usepackage{textcomp}%
\usepackage{manyfoot}%
\usepackage{booktabs}%
\usepackage{algorithm}%
\usepackage{algorithmicx}%
\usepackage{algpseudocode}%
\usepackage{listings}%
\usepackage{url}

\begin{document}
	
	\title{{A data-driven analysis on the mediation effect of compartment models between control measures and COVID-19 epidemics}}
	
	
	\author[1,2]{\fnm{Dongyan} \sur{Zhang}}\equalcont{These authors contributed equally to this work.}
	
	\author[3]{\fnm{Wuyue} \sur{Yang}}\equalcont{These authors contributed equally to this work.}
	
	\author[1]{\fnm{ Wanqi} \sur{Wen}}
	
	\author[4]{\fnm{Liangrong} \sur{Peng}}
	
	\author*[2]{\fnm{Changjing} \sur{Zhuge}}\email{zhuge@bjut.edu.cn}
	
	\author*[1]{\fnm{Liu} \sur{Hong}}\email{hongliu@sysu.edu.cn}

	\affil[1]{\orgdiv{School of Mathematics}, \orgname{Sun Yat-Sen University}, 
			\city{Guangzhou}, \state{Guangdong}, \postcode{510275}, \country{P. R. China}}
		
		\affil[2]{\orgdiv{Beijing Institute for Scientific and Engineering Computing, \\ Department of Mathematics, Faculty of Science}, \orgname{Beijing University of Technology}, 
				\city{Beijing}, \postcode{100124},  \country{P. R. China}}
			
			\affil[3]{\orgname{Yanqi Lake Beijing Institute of Mathematical Sciences and Applications},  \city{Beijing}, \postcode{101408}, \country{P. R. China}}
			
			\affil[4]{\orgdiv{College of Mathematics and Data Science}, \orgname{Minjiang University}, \orgaddress{ \city{Fuzhou}, \postcode{350108}, \state{Fujian}, \country{P. R. China}}}
			
			\abstract{\textbf{Background} \ On May 5th, 2023, WHO declared an end to the global COVID-19 public health emergency, which means a significant transition from global critical emergency response activities to long-term sustained COVID-19 prevention and control. At this very moment, we make a retrospective review on various control measures taken by 127 countries/territories during the first wave of COVID-19 pandemic until July 7, 2020, and evaluate their impacts on the epidemic dynamics quantitatively.
				
				\textbf{Methods} \  {We collect and evaluate the control measures implemented by each country. The SEIR-QD model, as a representative for general compartment models, is used to fit the epidemic data, enabling the extraction of crucial model parameters and dynamical features. The mediation effect of the SEIR-QD model is revealed by using the mediation analysis with structure equation modeling for multiple mediators operating in parallel. The inherent impacts of these control policies on the transmission dynamics of COVID-19 epidemics are clarified and compared with results derived from both multiple linear regression and neural-network-based nonlinear regression.}
				
				\textbf{Results} \ {A dramatic distinction in the control measures is observed among major countries/territories around the world, which largely affects the spreading rate and infected population size in each country. Several key dynamical features, like the normalized cumulative numbers of confirmed/cured/death cases on the 100th day and the half time, show statistically significant linear correlations with the control measures, thereby confirming the dramatic impacts of control measures and their respective implementation intensity on the epidemic. Most importantly, the SEIR-QD model, especially its infection rate and protection rate, has been confirmed to exhibit a statistical significant mediation effect between the control measures and dynamical features of epidemics. The mediation effect along the pathway from control measures in Category 2 to four dynamical features, through the infection rate, highlights the crucial role of nuclei acid testing and suspected cases tracing in effectively containing the spreading of epidemics.}
				
				\textbf{Conclusions} \ {Through a data-driven analysis, the mediation effect of compartment models is confirmed, which provides a better understanding on the intrinsic correlations among the strength of control measures and the dynamical features of COVID-19 epidemics.}}

			
			
			
			\maketitle
			
			\section{Background}\label{sec1}
			
			Since its first emergence at the end of 2019, the Corona Virus Disease 2019 (COVID-19) has swept across more than 222 countries and territories around the world. According to the statistics of the World Health Organization (WHO), more than 750 million confirmed cases of COVID-19 have been officially recorded worldwide till May 3rd, 2023, among which nearly 7 million people have died from this disease \cite{UNnews}. Though the real death number could be as large as 20 million, three times higher than the reported.
			
			During its spreading, the primitive novel coronavirus, SARS-CoV-2 has mutated into several variants, like Delta and Omicron (first identified in October 2020 and November 2021, respectively), which have higher transmission rates and ability of immune escape \cite{tian2022emergence}.
			In order to contain the pandemic, all countries/territories have adopted different levels of protective measures by controlling the infectious source, cutting off transmission routes and protecting susceptible populations. In addition to vaccination, measures like quarantining infected cases, tracing close contacts, travel restrictions, border closures, wearing face masks in public area and other physical distancing interventions, have played an important role in containing the pandemic.
			
			Recently, WHO declared an end to the global COVID-19 public health emergency on May 5th, 2023.  This means a significant transition from critical emergency response activities to long-term sustained COVID-19 disease prevention, control and management. Therefore, now it is a suitable moment to review various control measures and evaluate their impacts during the combat against COVID-19 pandemic.
			
			
			Actually, there have been a lot of cross-national comparative analyses on COVID-19 control measures \cite{balmford2020cross,Cascini:2022aa,Iyanda:2020aa,Balmford:2020aa}. For example, Hale et al. \cite{hale2021global} constructed an influential database, called the \emph{Oxford COVID-19 Government Response Tracker} (OxCGRT), to capture the government policies for over 180 countries/territories. With 19 policy indicators including containment and closure, economic response and health systems, OxCGRT enables further research that integrates the policy responses with epidemiological indicators.
			Analogous dataset can be found in Ref. \cite{cheng2020covid} too.
			Balmford et al. \cite{balmford2020cross} made cross-country comparisons and showed that policy interventions, rather than the socio-economic factors, determine the majority of variations in death rates of COVID-19 among OECD countries/territories.
			Unruh et al. \cite{unruh2022comparison} compared COVID-19 health policy responses in Canada, Ireland, UK and US, and concluded that the health system capacity, governance and political leadership all shaped country responses. To evaluate and compare the effectiveness of various policies, Koh et al. \cite{koh2020estimating} regressed some physical distancing measures, including international travel controls, restrictions on mass gatherings, and lock down, on time-varying reproduction number $R_t$. 
			
			Based on mathematical models, more quantitative analyses could be carried out. For example, sensitivity analysis of a dynamical model in \cite{tang2022lessons} showed that the stringent public health interventions are vital to contain the epidemic in China, while the high detection rate is vital to South Korea. As to municipal policy responses to COVID-19, Armstrong et al. \cite{armstrong2020measuring} used a survey in Canada to measure the aggressiveness of responses. This latent variable was found to be closely related to municipal population size and COVID-19 case. Zhou et al. \cite{zhou2022effectiveness} divided the population into 3 groups and extended the susceptible-exposed-infectious-hospitalized-removed model to adapt the dynamic zero-COVID policy in China. By examining the outbreaks of Delta variant in Xi'an, Yangzhou and Guangzhou in China, the key role of closed-off management, tracing and testing is highlighted.  
			
			{In this work, we restrain ourselves to the retrospective studies on the cross-national control measures taken during the first wave of COVID-19 and their impacts, since these measures and their relative implementation intensity during this time period are more diverse. The first aim of our work is to quantify the intrinsic close correlations between the control measures and the dynamical features of epidemics, which would help to reveal more efficient design and implementation of policies to contain analogous pandemics.
				Our second aim is to elucidate the significant roles of compartment models in the study of epidemics from an new perspective -- its mediation effect. This effect has seldom been mentioned in this filed but as we will show that it plays a critical  role in understanding the deep intrinsic connections between control measures and the transmission dynamics of COVID-19 epidemics.}
			
			This work is structured as follows. 
			{In Section \ref{sec2}, the epidemic data and control measures of 127 countries/territories are collected, a novel compartment model for epidemics is presented, the mediation analysis with structural equation modeling is discussed, and methods on clustering, correlation and regression analyses are briefly introduced. 
				Our main results including the classification of control measures, the clustering of 127 countries/territories, the  spreading dynamics of COVID-19 characterized through compartment models, impacts of control measures on the spreading dynamics and the mediation effect of compartment models are reported in Section \ref{sec3}.}
			The conclusion is presented in Section \ref{sec4}.

			\section{Methods}\label{sec2}
			\subsection*{Data collection}
			{To assess the influence of control measures on the transmission dynamics of COVID-19 in a quantitative way, here we collect 16 different types of control measures and their respective implementation strength marked by discrete grades from 127 countries/territories (see data file 1 in Supporting Information), which have been promulgated and taken by each government during the first wave of COVID-19 epidemics before July 7, 2020.} Although, strictly speaking, the last category -- medical resources is not a measure for prevention or containment, we still take it into consideration, because it characterizes the capacity of a country to provide professional medical care of those infected people and thus would influence the cure and mortality rates to a great extent. {On the other hand, three groups of official epidemic data, including numbers of daily new cases, cured cases and death cases, are collected from the WHO's reports \cite{policy16} (see data file 2 in Supporting Information).} For consistency, we redefine the outbreak date of COVID-19 epidemic in each country as the one when the number of accumulative cases in this country reaches 100. 
			
			\begin{table*}[t]
				\centering
				\renewcommand{\tablename}{\textsf{Table}}
				\includegraphics[width=1\linewidth]{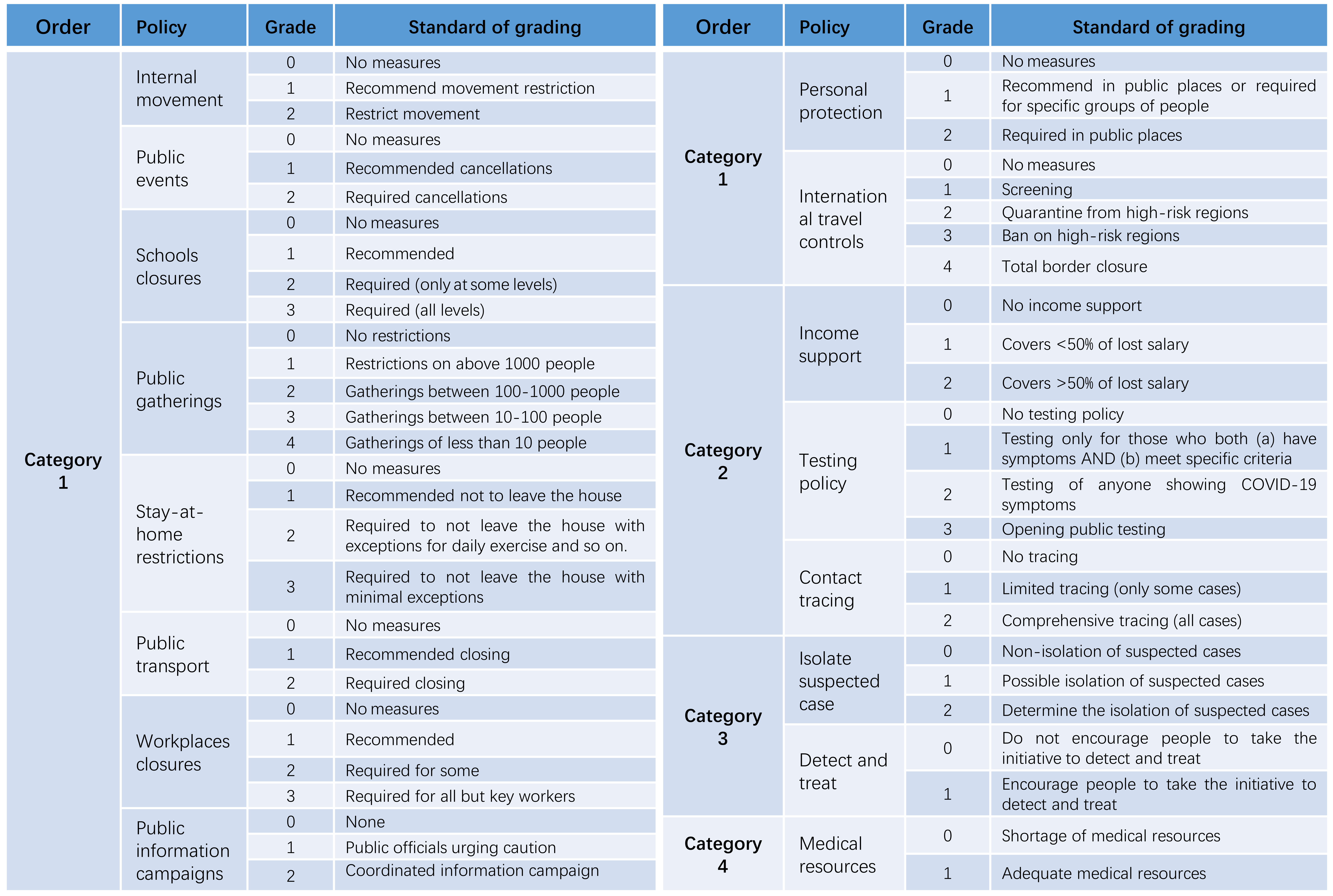}
				\caption{Summary of main policies taken by each country till July 7, 2020 during the first wave of COVID-19 pandemics. The strength of policies is characterized as ordinal data with 2-4 levels. Policy ``Schools closures'' in Category 1 is taken from \cite{acaps}; other policies in Category 1 and policy in Category 4 from \cite{owidcoronavirus}; policies in Category 2 from \cite{policy16}; policy ``Isolate suspected case'' in Category 3 from  \cite{policiy11-1,policy11-2,policy11-3,policy11-4}.}
				\label{tab:1}
			\end{table*}

			\subsection*{Mediation analysis with structural equation modeling}
			
			The mediation indicates that the effect of an independent variable on a dependent variable is transmitted through a third variable called mediator, which is frequently utilized in research of applied disciplines, such as psychology, education and social science \cite{baron1986moderator}. To go beyond single variable, one can refer to the Structural Equation Modeling (SEM) \cite{WenZhonglin2012}, which enables simultaneously analyzing multiple independent variables and multiple mediator variables, as well as dealing with manifest and latent variables at the same time. 
			
			For our purpose, here we illustrate the basic idea of structural equation modeling for mediation analysis with multiple mediators operating in parallel. Without loss of generality, let $X_1$ and $X_2$ be two independent variables, $Y$ the dependent variable, and $M_1$ and $M_2$ the mediators independent of each other. 
			The corresponding linear regression models read:
			\begin{subequations}
				\begin{align}
					Y&=\tau_{1}X_1 +\tau_{2}X_2 + \varepsilon_X,\\    
					M_1&=\eta_{11}X_1 +\eta_{12}X_2 + \varepsilon_1,\\     
					M_2&=\eta_{21}X_1 +\eta_{22}X_2 + \varepsilon_2,\\    
					Y&=\tau_{1}' X_1 +\tau_{2}'X_2 + \mu_{1}M_1+ \mu_{2}M_2+\varepsilon_Y.
				\end{align}
			\end{subequations}
			Here the first line regresses the dependent variable solely on independent variables, the second and third lines regress the mediators on independent variables, while the last line regresses the dependent variable on both the mediators and independent variables. 
			$\tau$'s and $\eta$'s are linear regression coefficients and $\varepsilon$'s denote errors. 
			
			Based on above formulas, the differences of coefficients 
			\begin{equation}
				\tau_{1}-\tau_{1}'= \mu_{1}\eta_{11}+ \mu_{2}\eta_{21}, \quad 
				\tau_{2}-\tau_{2}'= \mu_{1}\eta_{12}+ \mu_{2}\eta_{22},
			\end{equation}
			are recognized as the total mediation effect of $X_1$  (and $X_2$, respectively) on $Y$, where the constitutive term $\mu_{i}\eta_{ij}$ represents the corresponding individual mediation effect from $X_j$ to $Y$ through $M_i$ ($i,j=1,2$).

			To conduct mediation analysis, the first step is to centralize the data by subtracting mean values from the independent variables, dependent variables, and mediators, respectively. Subsequently, we employ the bootstrap technique \cite{WenZhonglin2012} to generate a resampled and sufficiently large dataset. 
			Finally, the SEM is used to estimate the undetermined coefficients.
			If the goodness of fit is acceptable, the significance of the mediation effect is judged based on its confidence interval obtained by the bias-corrected bootstrap. The criteria is, for the mediation effect under testing, if its confidence interval does not contain zero, the corresponding mediation effect is considered to be significant. Otherwise, if the fit of the SEM is unacceptable, the analysis will be stopped. 
			
			
			With respect to the compartment model in this study, the four categories of policies are considered as independent variables $(X_1, \cdots, X_4)$, whose indicators are 16 control measures in Table \ref{tab:1}. The logarithm of the five coefficients in the SEIR-QD model are used as the mediator variables, i.e. $(M_1,\cdots, M_5)=(\log(\alpha), \log(\beta), \log(\gamma), \log(\delta),
			\log(\kappa))$. Finally, the logarithms of six dynamical features characterizing the spreading of COVID-19 epidemics are taken as the dependent variable one by one, meaning $Y\in\{\log(Q_{100}/N), \log(R_{100}/N), \log(D_{100}/N), \log(t_{1/2}), \log(t_{lag}), \log(k_{app})\}$. The ``\textit{lavaan}'' package in R language \cite{lavaan2012} is used to perform our mediation analysis with structural equation modeling. The corresponding results can be found in data file 3 in Supporting Information.
			
			\color{black}
			\subsection*{Compartment models for epidemics}
			As a newly emerging highly infectious disease, COVID-19 has some unique characteristics in comparison to other infectious diseases, e.g., a relatively high ratio of asymptomatic patients. Besides, during the first wave of COVID-19 pandemics, public health emergency responses, such as locking down the domestic and international public transportation, quarantining infected patients and so on, had been implemented globally. These facts make the dynamic behaviors of COVID-19 epidemics dramatically different from previous ones. 
			
			By taking the specificity of COVID-19 into consideration, the SEIR-QD model (Eqs. \eqref{eq:ode:S}-\eqref{eq:ode:P}) has demonstrated its capability in modeling the transmission dynamics of COVID-19 epidemics and attracted extensive attentions and followers \cite{peng2020epidemic,yang2021rational}. Compared to the classic SEIR model, three additional compartments have been introduced into this model, including a compartment representing the protected individuals due to the strengthened prevention measures, a compartment for quarantined patients which cannot infect susceptible individuals as well as a separated recovery compartment from the one for death cases. 
			
			In summary, the SEIR-QD model contains seven compartments -- susceptible cases $S(t)$, protected and insusceptible cases $P(t)$, exposed cases $E(t)$, infected cases $I(t)$, quarantined cases $Q(t)$, recovered cases $R(t)$, and death cases $D(t)$. The transition rates between two connecting compartments are given through six parameters:  $\alpha$ characterizing how fast the susceptible cases convert into insusceptible cases, $\beta$ for the rate of susceptible cases getting infected upon contacting with infectious cases, $\gamma$ being the rate of exposed cases convert into infectious cases, $\delta$ as the rate of infectious cases into quarantined, while $\lambda$ and $ \kappa$ representing the rate of quarantined cases being recovered or dead. Moreover, $N$ denotes the total population number. 
			\begin{subequations}
				\begin{align}
					\dfrac{\mathrm{d}S(t)}{\mathrm{d}t} &= -\beta \dfrac{S(t)I(t)}{N} - \alpha S(t),\label{eq:ode:S}\\
					\dfrac{\mathrm{d}E(t)}{\mathrm{d}t} &= \beta \dfrac{S(t)I(t)}{N} - \gamma E(t),\label{eq:ode:E}\\
					\dfrac{\mathrm{d}I(t)}{\mathrm{d}t} &= \gamma E(t) - \delta I(t),\label{eq:ode:I}\\
					\dfrac{\mathrm{d}Q(t)}{\mathrm{d}t} &= \delta I(t) - \lambda Q(t) - \kappa Q(t),\label{eq:ode:Q}\\
					\dfrac{\mathrm{d}R(t)}{\mathrm{d}t} &= \lambda Q(t),\label{eq:ode:R}\\
					\dfrac{\mathrm{d}D(t)}{\mathrm{d}t} &= \kappa Q(t),\label{eq:ode:D}\\
					\dfrac{\mathrm{d}P(t)}{\mathrm{d}t} &= \alpha S(t).\label{eq:ode:P}
				\end{align}
			\end{subequations}
			
			By utilizing the simulated annealing algorithm in MATLAB 2021b \cite{MATLAB2021b}, the parameters in the SEIR-QD model are fitted with respect to the real epidemic data we collected. In this way, optimal values for the free model parameters are estimated automatically (see data file 4 in Supporting Information).

			\subsection*{Data clustering}
			Due to the significant diversity in the circumstances of each country/territory, it is neccessary to analyze the data for each class, which requires clustering the countries/territories before subsequent analysis. Here the K-means method is employed with the hyperparameter $4$ for the number of data clusters and Euclidean distance as the similarity metric for data samples.
			
			To test the significance of the difference among clusters, the Kruskal-Wallis rank sum test is carried out. The basic idea of Kruskal-Wallis test is to replace the original observation value with its rank for one-way analysis of variance. Suppose there are $M$ samples which are divided into $k$ groups. The sample size in the $i$-th group is $m_i$, and $M=\Sigma^k_{i=1}m_i$. Order all data samples from small to large and record the order number as their rank. In particular, if there are several data at the same position, take their average order number as the common rank. Once the value of Kruskal-Wallis $H$ test statistic
			\begin{equation}
				H = \dfrac{12}{M(M+1)}\sum_{i=1}^k\dfrac{R_i^2}{m_i} - 3(M+1) \sim \chi^2_{k-1}
			\end{equation}
			is larger than the significance level, where $R_i$ denotes the rank sums, one can conclude the sample groups are statistically significantly different.
			
			\subsection*{Correlation analysis}
			
			The Pearson's correlation coefficients are used to measure the linear correlation between random variables. Supposing two random variables $A$ and $B$ have $n$ scalar observed values, the Pearson's correlation coefficient of $A$ and $B$ is defined by
			\begin{equation}
				\rho(A,B) = \dfrac{1}{n-1}\sum_{i=1}^{n}(\dfrac{A_i-\mu_A}{\sigma_A})(\dfrac{B_i-\mu_B}{\sigma_B}),
			\end{equation}
			where $\mu_A$ and $\sigma_A$ are the mean and standard deviation of random variable $A$, $\mu_B$ and $\sigma_B$ are those of $B$ respectively. The calculation of Pearson's correlation coefficients can be directly extended to multiple random variables. Here we use the ``Corrcoef'' function of Matlab to perform the calculation and use the ``heatmap'' function for visualization.

			\subsection*{Multiple linear and nonlinear regressions}
			{To provide a relatively comprehensive description of the epidemic dynamics, we refer to six dynamic features, which are constituted by the cumulative numbers of confirmed/cured/dead cases on the 100th day normalized by the total population of the country ($Q_{100}/N$, $R_{100}/N$ and $D_{100}/N$), the half time $t_{1/2}$ (days required for reaching one half of the cumulative number of quarantined cases on the 100th day), the lag time $t_{lag}$ (days required for reaching $5\%$ of the cumulative number of quarantined cases on the 100th day), and the apparent spreading rate $k_{app}$ (the slope of the curve for the cumulative number of confirmed cases at the half time).}
			
			The influence of control measures on the transmission dynamics of COVID-19 epidemics is a key issue considered in the current study, and is analyzed by both linear and nonlinear regression models. Here, the multiple linear regression is performed mainly by the least squares method and is supplemented by the stepwise regression and the principal component regression by using the functions ``regress'', ``stepwise'' and ``pcacov'' in the Statistics and Machine Learning Toolbox of MATLAB \cite{MATLAB2021b}.
			
			The nonlinear regression analysis is realized through a neural network model -- a multilayer perceptron (MLP) constituted by one input layer, one hidden layer with 100 nodes and the ReLU activation function $\sigma(z)=\max (0, z)$, and one output layer using the Sigmoid activation function $\sigma(z)=(1+e^{-z})^{-1}$. The inputs are normalized grades of 16 control measures, while the outputs are values of six dynamical features for the epidemic. The $L_2$ norm of the deviations between true values and the predicted ones gives the loss function. The network is trained for 10000 epochs with the Adam optimizer and the learning rate is set to $10^{-3}$. In addition, reasonably adjusting the batch size of the whole dataset will accelerate convergence, which in this case is set as $10$.
			
			\section{Results}\label{sec3}
			\subsection*{Control measures of different countries are highly diverse during the first wave of the COVID-19 pandemic}
			
			As the 16 different types of control measures for 127 countries/territories involved in this study (see \textit{Methods} and Table \ref{tab:1}) are not mutually independent, e.g. strict restrictions on public gathering would be closely related to the measures of school and workplace closure, it is necessary to analyze the correlations among them. 
			
\begin{figure*}[t]
				\centering
				\includegraphics[width=1\linewidth]{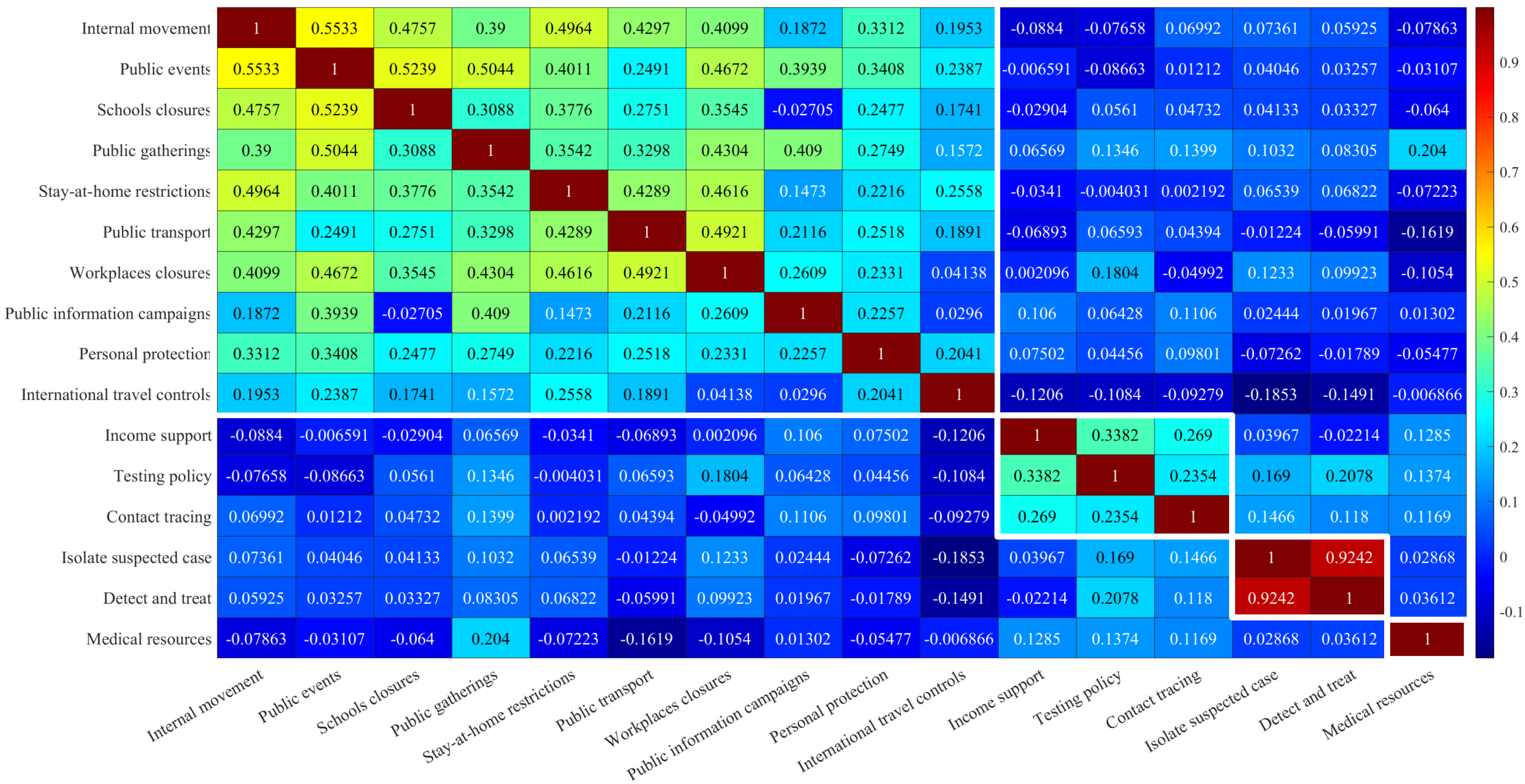}
				\renewcommand{\figurename}{\textsf{Fig.}}
				\caption{Heatmap for correlations among control measures of 127 countries/territories. Dark red indicates a strong positive correlation close to 1 between the two measures labelled by horizontal and vertical axes, while dark blue means a weak correlation.}
				\label{fig:heatmap}
			\end{figure*}

			To this end, heatmap is used for descriptive analysis of the control measures (Fig. \ref{fig:heatmap}), according to which, it is clear that the sixteen control measures can be aggregated into four categories (Table \ref{tab:1}). The first category includes restrictions on public transportation and closure of public places, which reduce the susceptible population and contacts among people. The second category includes policies on contact tracking and testing, which directly influence the extent and efficacy of case detection and documentation. 
			The third category refers to the promotion of testing and the implementation of quarantine measures on suspected cases, reflecting the efficacy of quarantine protocols.
			The last category includes the status of medical resources.
			
\begin{figure*}[htp]
				\centering
				\includegraphics[width=1\linewidth]{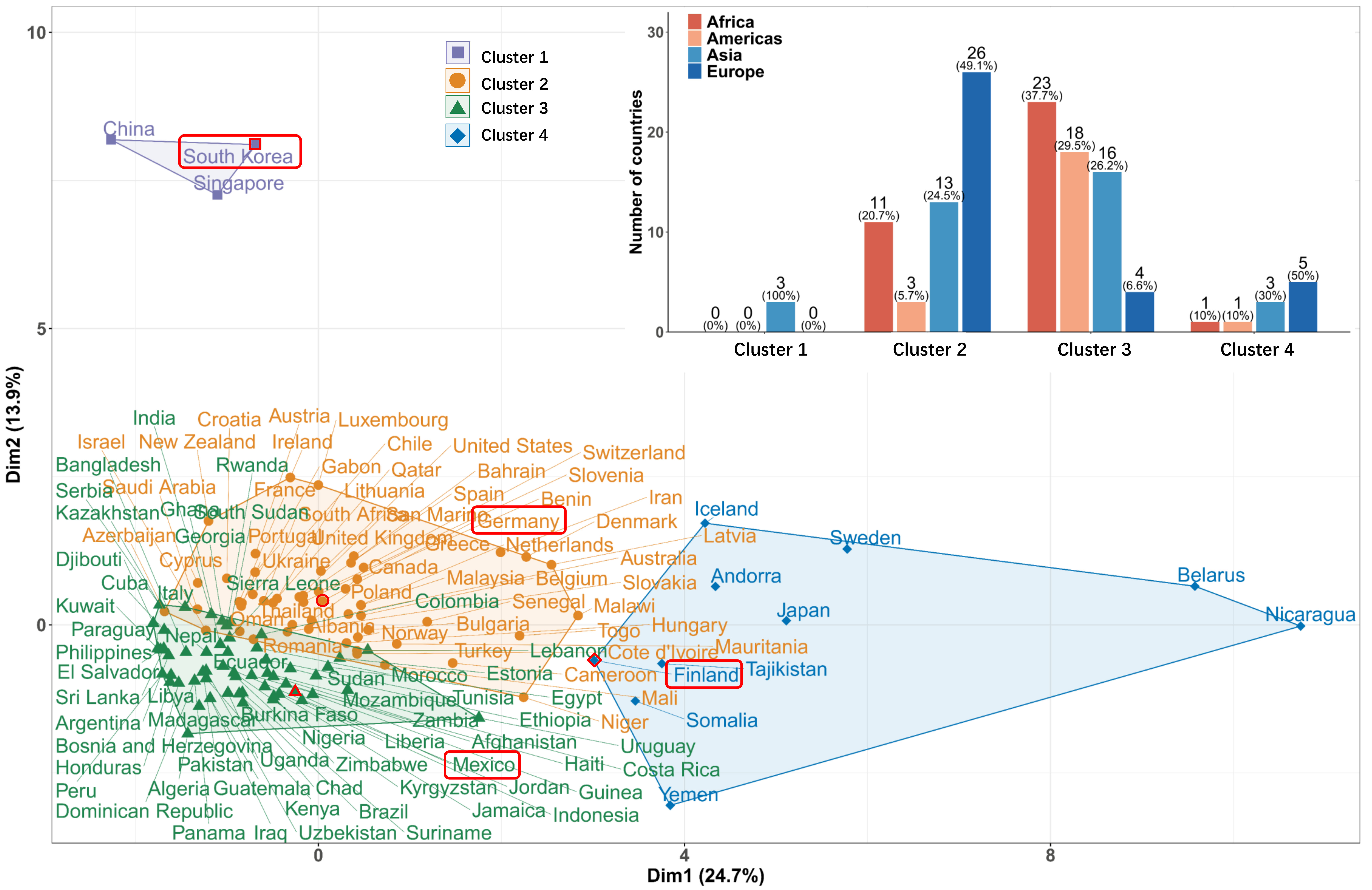}
				\renewcommand{\figurename}{\textsf{Fig.}}
				\caption{Clustering analysis of countries/territories by K-means algorithm based on their implementation intensity of control measures. The t-SNE method is used for data visualization. The distribution among the continents of each group is shown in the inset.}
				\label{fig:1}
			\end{figure*}

			On the other hand, it has been widely known that there is a dramatic distinction among the control measures and their implementation intensity taken by each nation during the first wave of COVID-19 pandemic. By using the K-means method, these countries are also clustered into 4 groups based on their own implementation strength of control measures. As visualized in Fig. \ref{fig:1}, \emph{\textbf{Cluster 1}} includes 3 Asian countries, China, South Korea, and Singapore, which takes strict control measures and keeps the pandemic well contained during the early 2020. \emph{\textbf{Cluster 2}} includes 53 countries/territories, more than half of which are European countries. Most countries in this group have rich medical resources and take mediate-level measures of prevention and containment. \emph{\textbf{Cluster 3}} comprises 61 nations/territories, predominantly encompassing developing countries of the Third World. These countries face inadequate medical resources, and the implementation of COVID-19 control measures within this cluster lacks the necessary stringency to yield desired outcomes. \emph{\textbf{Cluster 4}} includes the remaining 10 countries. This cluster is of greater diversity than the preceding three ones. Similar results are obtained by the agglomerative hierarchical clustering method and the density-based spatial clustering and application with noise method, (see Figs. S1, S2 in Supporting Information), which confirm the robustness and reliability of our clustering.
			
			\subsection*{Control measures and their implementation strength have significant impacts on the epidemic dynamics}
			
			{To provide a comprehensive description of the transmission dynamics of the first wave of COVID-19 epidemics, six dynamical features are introduced, that is $Q_{100}/N$, $R_{100}/N$, $D_{100}/N$, $t_{1/2}$, $t_{lag}$, and $k_{app}$. Their definitions are given in the section of Methods, and their values have been extracted from the WHO data (see data file 4 in Supporting Information).}
			
			As illustrated in Fig. \ref{fig:BoxPlots}A, the half time $t_{1/2}$ and the apparent spreading rate $k_{app}$ show a clear monotonic dependence on the control strength over the countries in the first three clusters. 
			The non-monotonic dependence for the cumulative confirmed, cured and death cases on the 100th day could be attributed to the fact the epidemics in those less developed countries/territories are still out of control on the 100th day since its outbreak. The extraordinarily long lag-time for countries in Cluster 1 shows that the spreading of COVID-19 virus has been suppressed at a very low level due to the extremely strict containing measures taken by these three countries.

			A more quantitative analysis is carried out by adopting multiple linear regression. Based on p-values and R-square values summarized in Figs. \ref{fig:BoxPlots}B-\ref{fig:BoxPlots}G, some general conclusions are reached, i.e.
			\begin{itemize}
				\item
				Except for the lag time, all other five dynamical features are statistically significantly correlated with the overall implementation intensity of control measures in each country, where the p-values for correlation coefficients are all less than $0.01$.
				\item
				The impacts of the control measures in Category 2 on the cumulative confirmed/cured/death cases on the 100th day $Q_{100}/N, R_{100}/N, D_{100}/N$ and the spreading rate $k_{app}$ are statistically significant. These findings reveal the importance of keeping track of suspected cases, who have had close contact with confirmed cases, for maintaining the epidemic under control.
				\item
				The p values for the linear correlations between the half time/lag time and measures in Category 4, the half-time and measures in Category 3 are less than $0.05$, meanwhile their corresponding R-square values are relatively small too. This fact reveals the intrinsic diversity and complexity of the spreading dynamics of COVID-19 epidemics. Solely raising up the control strength of single (or several) measure may not be as effective as expected.	
				\item 
				With respect to nations in Cluster 2, statistically significant correlations are observed between $Q_{100}/N$ and measures in Category 1, $D_{100}/N$ and measures in Category 1 and Category 4. Meanwhile, as a contrary, correlations are absent for nations in Cluster 3 (with p-values much larger than $0.05$). This result hints that strengthening the public control, like school and workplace closures, can be effective only when the epidemic is still under control. So is the increase of medical resources for reducing the death cases.
			\end{itemize}
			
			\begin{figure*}[htpb]
				\centering
				\includegraphics[width=0.9\linewidth]{fig3.png}
				\renewcommand{\figurename}{\textsf{Fig.}}
				\caption{Impacts of control measures on the spreading dynamics of COVID-19 epidemics. (A) Variations of 6 epidemic dynamical features among countries belonging to different clusters are illustrated through box plots. After taking base-10 logarithmic transformation, their linear correlations with control measures in different categories for (B-C) all nations or nations in (D-E) Cluster 2 and (F-G) Cluster 3 are highlighted through (B,D,F) the p-values (the logarithm to be exact) and (C,E,G) R-square values separately. Notice the p-values and R-square values are not applicable to Category 3 for nations in Cluster 3 in (F) and (G).}
				\renewcommand{\figurename}{Fig.}
				\label{fig:BoxPlots}
			\end{figure*}
			
			\color{black}
			\subsection*{The SEIR-QD model exhibits a statistically significant mediation effect}

			In the last section, we have shown that the control measures exhibit statistically significant impacts on the spreading dynamics of COVID-19 epidemics. However, to gain a better understanding on their intrinsic correlations, especially to aid public healthy policy decision, we need to refer to concrete predictive models. Here we make use of the SEIR-QD model in Eqs. \eqref{eq:ode:S}-\eqref{eq:ode:P}, which has been shown to be a suitable model for studying COVID-19 epidemics  \cite{peng2020epidemic,yang2021rational}. 
			
\begin{figure*}[htp]
				\centering
				\includegraphics[width=1\linewidth]{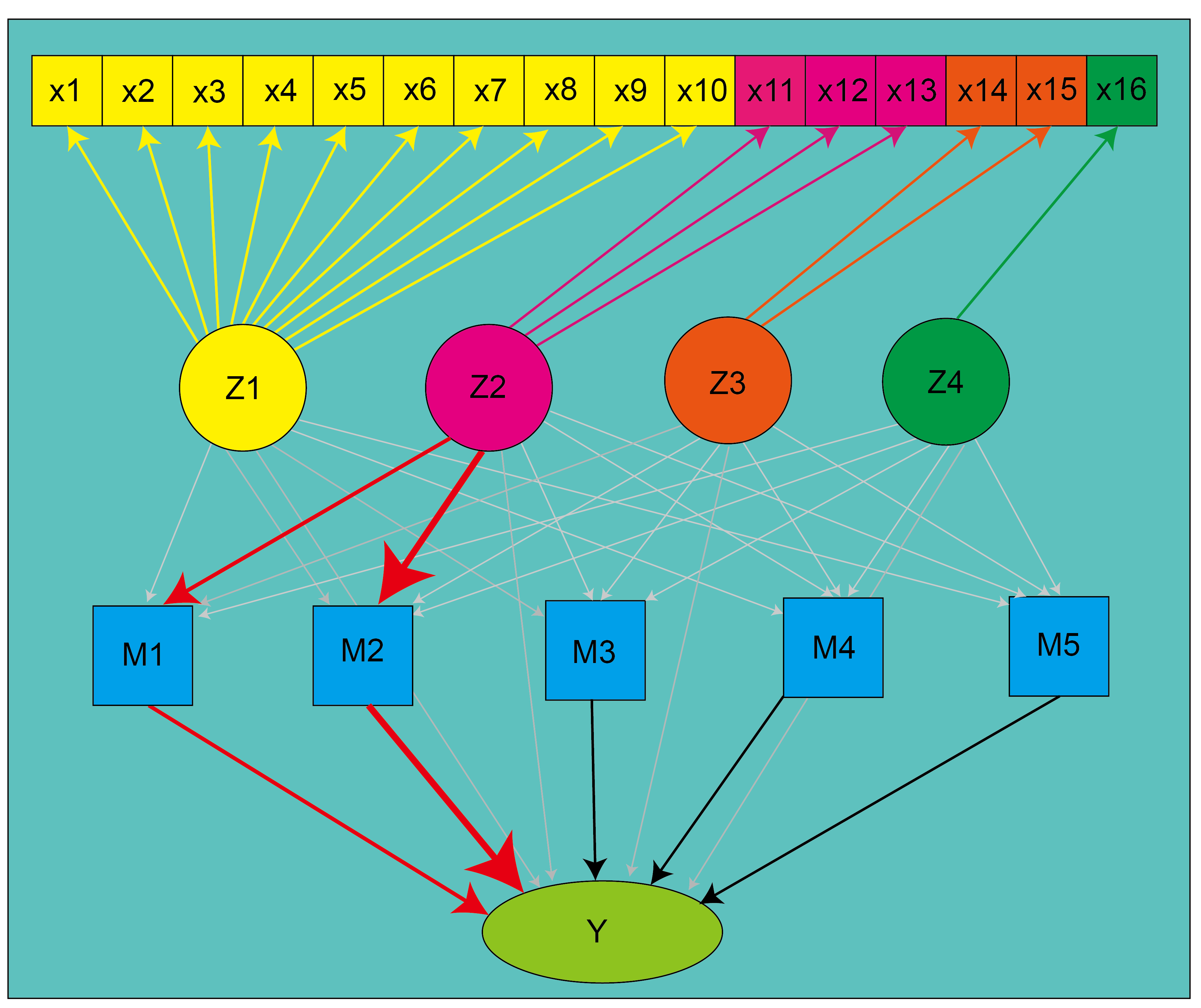}
				\renewcommand{\figurename}{\textsf{Fig.}}
				\caption{The mediation effect of SEIR-QD model between control measures and COVID-19 epidemic dynamics. $x_1-x_{16}$ denote 16 control measures summarized in Table \ref{tab:1}, which have been clustered into four categories $Z_1,Z_2,Z_3,Z_4$. $M_1-M_5$ represent the logarithm of five coefficients in the SEIR-QD model, and the scalar $Y$ takes each of the 6 dependent variables, $\{\log(Q_{100}/N), \log(R_{100}/N), \log(D_{100}/N), \log(t_{1/2}), \log(t_{lag}), \log(k_{app})\}$, in sequence. The pathways exhibiting  statistically significant mediation effects are highlighted.}
				\renewcommand{\figurename}{Fig.}
				\label{fig:mediation}
			\end{figure*}

			Based on the mediation analysis with structural equation modeling (see \textit{Methods}) \cite{WenZhonglin2012, lavaan2012}, the SEIR-QD model is found to play a crucial mediation effect between control measures and the epidemic dynamics. To be concrete,
			\begin{itemize}
				\item
				The total mediation effect of the SEIR-QD model is statistically significant for four epidemic dynamical features, i.e. the normalized cumulative numbers of confirmed/cured/death cases on the 100th day $Q_{100}/N, R_{100}/N, D_{100}/N$ and the spreading rate $k_{app}$.
				\item
				The infection rate $\beta$ plays a significant role in the mediation effect of the SEIR-QD model. In particular, the mediation effect along the pathway from control measures in Category 2 to the four dynamical features through the infection rate $\beta$ is the most prominent, as illustrated through red arrows in Fig. \ref{fig:mediation}. This fact can be clearly understood, since more frequent nucleic acid tests and continued tracking of suspected cases allow for a shorter time to identify these infected cases and isolate them from normal people, and thus effectively reduce the infection rate $\beta$.
				\item The mediation effect along the pathway from control measures in Category 2 to $Q_{100}/N$ through the protection rate $\alpha$ is also statistically significant. It shows that the inclusion of the protection rate $\alpha$ into the classical SEIR model is crucial for proper characterization of the impacts of control measures on the COVID-19 epidemics.
			\end{itemize}
			Our above findings agree with the report by Zhou et al. \cite{zhou2022effectiveness}, in which they also highlighted the key role of tracing and testing during the outbreaks of Delta variant in Xi’an, Yangzhou and Guangzhou in China.
			
			\color{black}

			\subsection*{The SEIR-QD model allows a robust quantitative evaluation on the impacts of control measures}
			
			The SEIR-QD model provides a quantitative way to estimate the impacts of various control measures and their implementation intensities on the transmission dynamics of COVID-19 epidemics. To begin with, we argue that there is a close relation between the control measures and parameters in the SEIR-QD model.
			\begin{itemize}
				\item
				Control measures in Category 1 mainly affects the protection rate $\alpha$ and infection rate $\beta$ in the SEIR-QD model. Increasing the control strength of policies in Category 1 will lead to a larger $\alpha$ and a smaller $\beta$.
				\item
				In addition to the infection rate $\beta$, the control measures of Category 2 adjust the quarantine rate $\delta$ too. More frequent nuclei acid testing and continuous tracking of suspected cases allow a shorter time to identify those infected cases and isolate them from normal people. This statement is directly supported by the linear correlation analysis, which shows that the correlations for both $\beta$ and $\delta$ are statistically significant, as well as the mediation analysis, which gives an affirmative answer to the coefficient $\beta$.
				\item
				The linear correlations between control measures in Category 3 \& 4 and the mortality rate $\kappa$ are statistically significant. Richer medical resources mean patients can get more professional care and medical treatment in general, resulting in a decrease in the mortablity rate $\kappa$.
			\end{itemize}
			
\begin{figure*}[htp]
				\centering
				\includegraphics[width=0.94\linewidth]{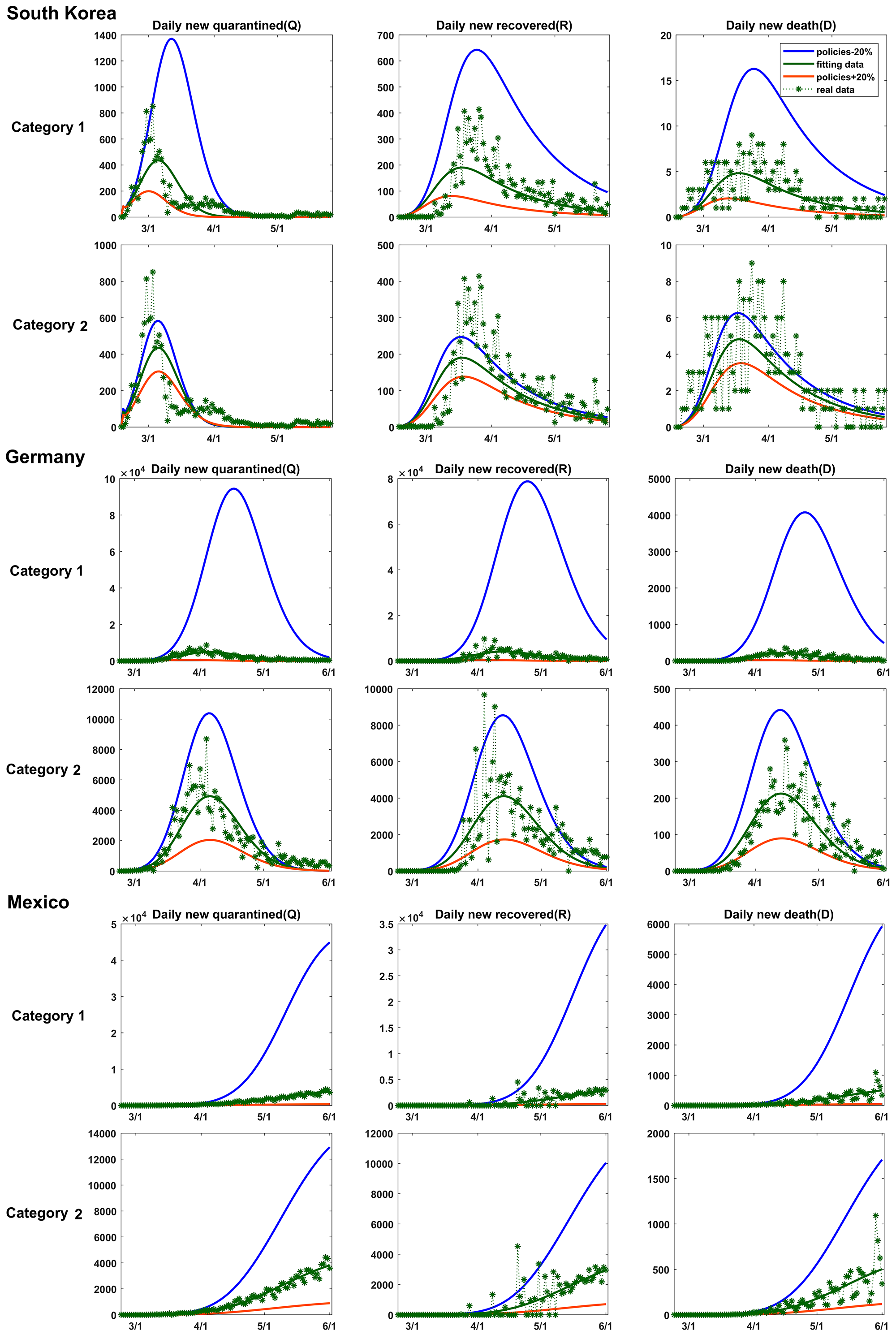}
				\renewcommand{\figurename}{\textsf{Fig.}}
				\caption{Impacts of control measures in Category 1 \& 2 on the COVID-19 epidemics evaluated through the SEIR-QD model. In comparison with base lines (real data: green stars with dashed lines, model fitting: green solid lines), parameters $\alpha$ and $\beta$ are changed by $+20\%$ (blue lines) and $-20\%$ (red lines) to simulate  the influence of control measures in Category 1. So are the parameters $\beta$ and $\delta$ for measures in Category 2. Meanwhile, all other parameters remain unchanged at their default values.}
				\renewcommand{\figurename}{Fig.}
				\label{fig:dynamic_model}
			\end{figure*}

			{Now based on above statements, the parameters corresponding to control measures in each category are adjusted by $\pm 20\%$ with respect to their default values, while the rest parameters in the SEIR-QD model are kept unchanged to reflect the different policy impacts. As depicted in Fig. \ref{fig:dynamic_model} through three representative countries -- South Korea for Cluster 1, Germany for Cluster 2, and Mexico for Cluster 3, the COVID-19 epidemics in countries belonging to different clusters exhibit significantly distinguishable reliance on control measures. For instance, a twenty percentage relaxation on the implementation intensity of measures in either Category 1 or Category 2 results in limited perturbations on the daily new quarantined cases in South Korea, whereas in Mexico, the impact becomes exceptionally large. Germany falls somewhere in between. The influence of control measures in Category 2 is weaker compared to that of measures in Category 1. These observations accurately reflect the varying situations in these countries. Interested readers are referred to the Supporting Information for data on other representative countries for further comparison and validation.}
			
\begin{figure*}[htp]
				\centering
				\includegraphics[width=1\linewidth]{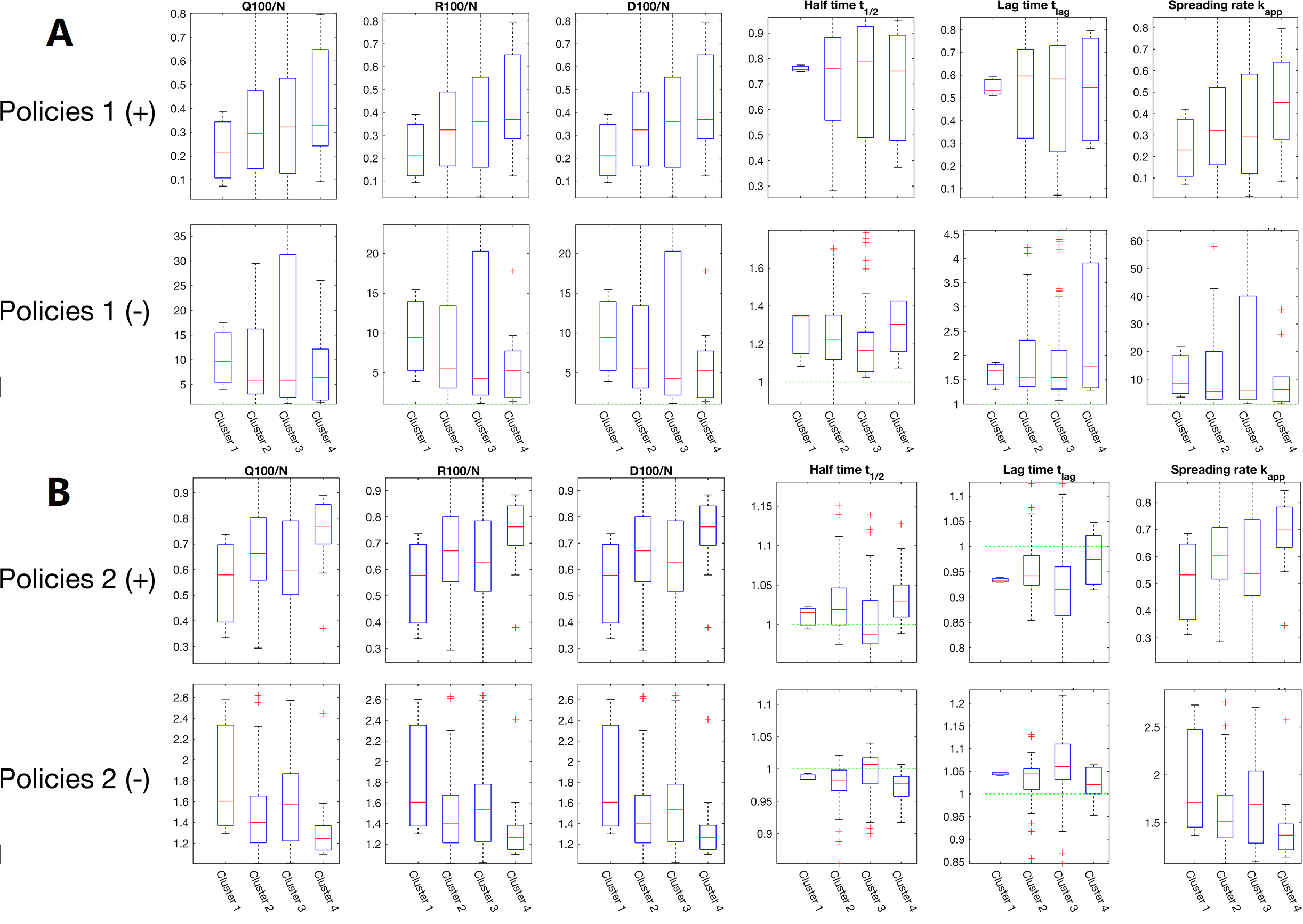}
				\renewcommand{\figurename}{\textsf{Fig.}}
				\caption{Impacts of control measures on the COVID-19 epidemics evaluated through the SEIR-QD model.  (A) Parameters $\alpha$ and $\beta$ are changed by $\pm 20\%$ to mimic the effects of measures in Category 1. Meanwhile, all other parameters are kept at their default values. For comparison, countries in four clusters are illustrated separately. Their correponding fold changes are indicated through the vertical axes. So are the parameters $\beta$ and $\delta$ for Category 2 in (B). }
				\renewcommand{\figurename}{Fig.}
				\label{fig:sensitivity}
			\end{figure*}

			{In addition, a global sensitivity analysis is performed and summarized in Fig. \ref{fig:sensitivity}. As measured by the normalized cumulative numbers of confirmed/cured/death cases on the 100th day, $Q_{100}/N, R_{100}/N, D_{100}/N$, tightening measures in Category 1 seems to be more effective than those in Category 2. However, the cost for shutting down public transportations and public places is much higher too. Furthermore, by increasing the control strength of measures in Category 1, the fold changes in $Q_{100}/N$ among four country clusters show a monotonic behavior, meaning its average impact on Cluster 1 is the strongest while on Cluster 3 and 4 are the weakest. Similar conclusions could be reached for $R_{100}/N$ and $D_{100}/N$ too. Contrarily, non-monotonic behaviors are observed for the impacts of control measures in Category 2 among four country clusters. The average influence of nuclei acid testing and keeping tracing suspected cases for countries in Cluster 3 is more apparent than expected. Especially, control measures in Category 1 and 2 exhibit opposite impacts on the half time. This fact shows that performing nuclei acid testing and keeping tracing suspected cases only suppress the outbreak of COVID-19 epidemics and thus prolong the time to reach its peak.  In summary, the policy impacts on the transmission dynamics of COVID-19 are significantly different among four country clusters.}

			{Along with above compartment-model-based analysis, the multiple linear regression and nonlinear regression by MLP are carried out too. The results by MLP show an excellent fitting to the real epidemic data, while the performance of a multiple linear regression model is moderate. However, neither linear nor nonlinear regression models offer consistent results during the sensitivity analysis (Figs. S4, S10, S11 in Supporting Information), and thus are not suitable for evaluating the policy impacts on the epidemic dynamics.}

			\section*{Discussion}
			
			In the field of epidemiology, a crucial issue is to what extent a country's implemented control measures and their respective strength can influence its own epidemic trend. It has sparked significant controversy, particularly during each epidemic caused by infectious diseases such as influenza, SARS, Ebola, and many others.
			In this paper, by performing comprehensive analyses based on the officially reported epidemic data of 127 countries/territories and 16 concrete control measures taken by each country during the first wave of COVID-19 epidemic, we obtain several preliminary yet insightful results.
			
			Apparently, our current study is far from complete. Firstly, the COVID-19 epidemic is an exceedingly complex process, which means numerous other factors may come into play alongside control policies. For example, the movement of people across borders, the presence of asymptomatic cases, the emergence of new mutant types of coronavirus and so on, make it almost impossible to keep a country free from infection, no matter what kinds of control measures are implemented. Furthermore, even with a strong willing, many countries/territories are facing with significant challenges in implementing high-level controls on the public gathering, workplace closure, etc., due to both economic and social issues. These facts are totally ignored in the current study to a great pity. We also constrain our analyses on the first-wave data of COVID-19 epidemic, which means the more fruitful phenomena appeared in the second and following waves of COVID-19, the temporal changes in the control measures of each country in response to the progressing of epidemics have not been taken into consideration yet.
			
			It is also a critical issue whether the results we obtained in the current article are robust and reliable. To avoid over-interpretation, the data are aggregated, the policies are grouped and the countries under study are clustered. Almost all statements are reached in the statistical sense over plenty of countries with similar situations. Meanwhile, uncertainty quantification on the accuracy and robustness of model parameters, as well as their consequence on the predicted epidemic dynamics have been carried out too (see Figs. S6, S7, S8 in the Supporting Information). Our preliminary results show that there is no apparent contradiction to the general conclusions in this work.
			\color{black}
			
			\section{Conclusion}\label{sec4}

			This paper elucidates the remarkable impacts of control measures and their effectiveness in implementation of each country on their own transmission dynamics of COVID-19 epidemics quantitatively. In particular, the mediation effect of compartment models, represented by the SEIR-QD model in the current study, is analyzed with respect to real epidemic data and proved to play a statistically significant role. This finding confirms the importance of compartment models during the study of epidemics from a new perspective.
			
			To help policy makers and researchers implement more effective prevention strategies tailored to their respective national contexts, we summarize the key findings of our study below. 
			\begin{itemize}
				\item
				During the first wave of COVID-19 pandemic, there is a dramatic distinction in the control measures and their implementation strength among major countries/territories in the world, which largely affects the spreading speed and infected population size in each country.
				\item 
				Several key dynamical features, like the normalized cumulative numbers of confirmed/cured/death cases on the 100th day, the half time and the apparent spreading rate, show statistically significant linear correlations with the overall control measures.
				\item
				The SEIR-QD model, especially the infection rate $\beta$ and protection rate $\alpha$, exhibits a statistically significant mediation effect between the control measures and dynamical features for epidemics. In particular, the mediation effect along the pathway from control measures in Category 2 to four dynamical features -- $Q_{100}, R_{100}, D_{100}$ and $k_{app}$ through the infection rate $\beta$ is the most prominent. This fact highlights the importance of nuclei acid testing and keeping tracing suspected cases to contain the epidemics under control.
				\item
				The compartment models, in particular the SEIR-QD model in the current study, allow a robust quantitative evaluation on the policy impacts. In contrast, no consistent result could be reached by either the multiple linear regression or neural-network-based nonlinear regression.  
			\end{itemize}
			\color{black}
			
			\subsection*{Abbreviations}  
			{\scriptsize
				\begin{tabular}{ll}
					COVID-19 & Corona Virus Disease 2019 \\
					MLP & MultiLayer Perceptron \\
					OxCGRT & Oxford COVID-19 Government Response Tracker \\
					SEM & Structural Equation Modeling \\
					WHO & World Health Organization
				\end{tabular}
			}
			\backmatter

			\subsection*{Supplementary information}
			
			The online version contains supplementary material available at https://doi.org/xxxxxxxxxxx.
			
			
			
			
			

			\section*{Declarations} \scriptsize
			
			\subsection*{Ethics approval and consent to participate}
			Not applicable.
			
			\subsection*{Consent for publication}
			Not applicable.
			
			\subsection*{Availability of data and materials} 
			The datasets generated during and/or analyzed in this study are available from the corresponding authors on reasonable request.

			\subsection*{Competing interests}
			The authors declare that they have no confict of interest.
			
			\subsection*{Funding}
			This work was supported by the National Natural Science Foundation of China (11801020, 12205135, 21877070), Guangdong Basic and Applied Basic Research Foundation (2023A1515010157), the Natural Science Foundation of Fujian Province of China (2020J05172),  Startup Research Funding of Minjiang University (mjy19033), Start-up Research Fund of Yanqi Lake Beijing Institute of Mathematical Sciences and Applications, Special Pre-research Project of Beijing University of Technology for Fighting the Outbreak of Epidemics.
			
			\subsection*{Authors' contributions} 
			LH and CZ designed and implemented the study, collected the data, analyzed the data, writing--original draft, writing--review \& edit. DZ, WW and WY analyzed the data, writing--original draft, writing--review \& edit. LP writing--original draft, writing--review \& edit. All authors have read and approved the content of the manuscript. All authors read and approved the final manuscript.
			
			\subsection*{Acknowledgements} 
			Not applicable.

			\bibliography{ref-1}

		\end{document}